\documentclass[english,prl,twocolumn,aps,amssymb,footinbib,showpacs]{revtex4}
\usepackage[T1]{fontenc}
\usepackage{amsmath}
\usepackage{amssymb}
\usepackage{graphicx}
\usepackage{braket}
\usepackage{babel}
\usepackage{color}
\usepackage{siunitx}

\makeatletter

\usepackage{braket}

\@ifundefined{textcolor}{}
{
 \definecolor{BLACK}{gray}{0}
 \definecolor{WHITE}{gray}{1}
 \definecolor{RED}{rgb}{1,0,0}
 \definecolor{GREEN}{rgb}{0,1,0}
 \definecolor{BLUE}{rgb}{0,0,1}
 \definecolor{CYAN}{cmyk}{1,0,0,0}
 \definecolor{MAGENTA}{cmyk}{0,1,0,0}
 \definecolor{YELLOW}{cmyk}{0,0,1,0}
 }

\usepackage{bm}\@ifundefined{definecolor}{\usepackage{color}}{}

\usepackage[T1]{fontenc}
\usepackage[latin9]{inputenc}
\usepackage{textcomp}
\usepackage{color}

\usepackage{babel}

\makeatother

\begin{document}

\title{Detecting spins with a microwave photon counter}

\author{ Emanuele Albertinale$^{1}$, L\'eo Balembois$^{1}$, Eric Billaud$^{1}$, Vishal Ranjan$^{2}$, Daniel Flanigan$^{1}$, Thomas Schenkel$^{3}$, Daniel Est\`eve$^{1}$, Denis Vion$^{1}$, Patrice Bertet$^{1}$, Emmanuel Flurin$^{1}$}
\email{emmanuel.flurin@cea.fr}
\affiliation{$^1$Universit\'e Paris-Saclay, CEA, CNRS, SPEC, 91191 Gif-sur-Yvette Cedex, France\\
$^2$National Physical Laboratory, Hampton Road, Teddington, Middlesex, TW11 0LW, UK \\
$^3$Accelerator Technology and Applied Physics Division, Lawrence Berkeley National Laboratory, Berkeley, California 94720, USA
}

\date{\today}

\begin{abstract}
Quantum emitters respond to resonant illumination by radiating electromagnetic fields. A component of these fields is phase-coherent with the driving tone, while another one is incoherent, consisting of spontaneously emitted photons and forming the fluorescence signal. Atoms and molecules are routinely detected by their fluorescence at optical frequencies, with important applications in quantum technology~\cite{kimble_photon_1977,wehner_quantum_2018} and microscopy~\cite{orrit_single_1990,klar_fluorescence_2000,betzig_imaging_2006,bruschini_single-photon_2019}. Spins, on the other hand, are usually detected by {their coherent response} at radio- or microwave frequencies, either in continuous-wave or pulsed magnetic resonance~\cite{schweiger_principles_2001}. Indeed, fluorescence detection of spins is hampered {by their low spontaneous emission rate} and by the lack of single-photon detectors in this frequency range. Here, using superconducting quantum devices, we demonstrate the detection of a small ensemble of donor spins in silicon by their fluorescence at microwave frequency and millikelvin temperatures. We enhance the spin radiative decay rate by coupling them to a high-quality-factor and small-mode-volume superconducting resonator~\cite{bienfait_controlling_2016}, and we connect the device output to a newly-developed microwave single photon counter~\cite{lescanne_irreversible_2020} based on a superconducting qubit. We discuss the potential of fluorescence detection as a novel method for magnetic resonance spectroscopy of small numbers of spins. 
\end{abstract}

\maketitle

\begin{figure*}
\includegraphics[width=1\textwidth]{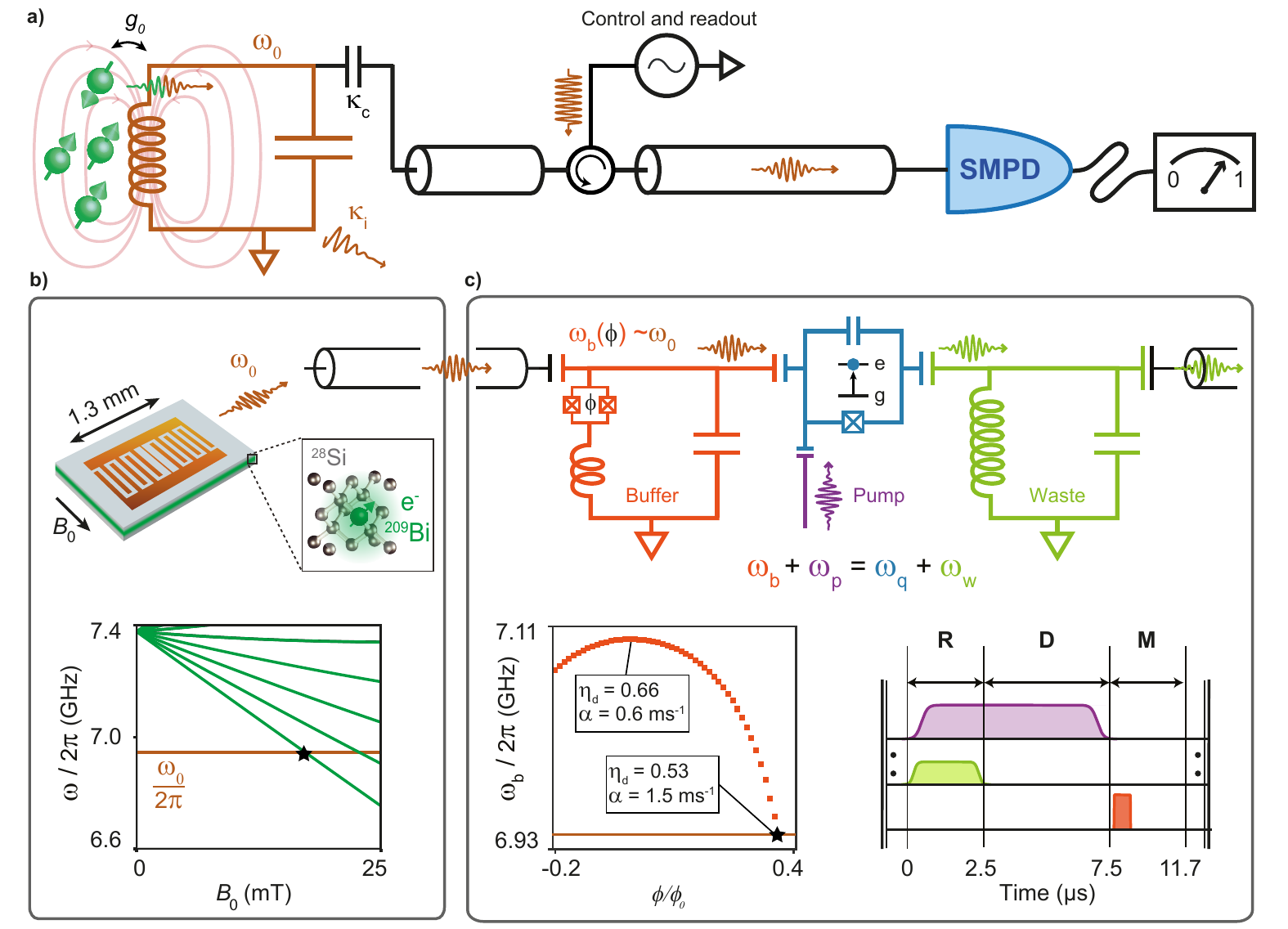}
 \caption{\textbf{Principle of spin detection with a photon counter.} 
 (\textbf{a}) Schematics of the experiment. Each spin in the ensemble is coupled with a strength $g_0$ to a resonator of angular frequency $\omega_0$ and internal loss rate $\kappa_i$, itself coupled with a rate $\kappa_c$ to an input-output line. This line allows to drive the spins with microwave pulses through a circulator and to route the photons emitted by the spins while relaxing radiatively towards a Single Microwave Photon Detector (SMPD).
 (\textbf{b}) Spin device schematics. The spins are bismuth donors implanted in a silicon substrate isotopically enriched in the $^{28}\mathrm{Si}$ isotope, on top of which a superconducting LC resonator is patterned. A magnetic field $B_0$ is applied parallel to the substrate to tune the lowest bismuth donor frequency in resonance with $\omega_0$ at $B_0 \sim17$\,mT. 
 (\textbf{c}) SMPD device and operation schematics. The SMPD relies on a transmon qubit at frequency $\omega_q / 2\pi = \SI{6.13}{GHz}$, coupled to three ports, namely buffer, pump, and waste. The buffer port consists of a resonator at frequency $\omega_b (\phi)$ that can be tuned to $\omega_0$ by applying a flux $\phi$ to a SQUID loop inserted in it (see inset). The waste port consists of a resonator at a fixed frequency $\omega_w/2\pi = \SI{7.63}{GHz}$. Incoming photons at the buffer are converted into excitation of the transmon qubit and into a photon in the waste resonator by a four-wave mixing process enabled by a pump tone sent via the pump port at frequency $\mathrm{\omega_p=\omega_q+\omega_w-\omega_b}$. The SMPD operation consists of a three-step cycle of total duration \SI{11.7}{\micro\second} which can be repeated continuously. The Reset (R) step is achieved by driving the waste port and the pump. The Detection step (D) is achieved by switching off the waste port, and keeping the pump on. The Measurement step (M) is performed by switching off the pump, and sending a microwave pulse on the buffer port to perform dispersive qubit readout. The SMPD efficiency $\eta_d$ and dark count rate $\alpha$ are indicated in the inset for relevant values of $\phi$ in the inset.   
 }
\label{fig1}
\end{figure*}

Microwave measurements at the quantum limit have recently become possible thanks to the development of superconducting parametric amplifiers that linearly amplify a signal at cryogenic temperatures with minimal added noise~\cite{roy_introduction_2016}. These advances enable efficient measurement of the field quadratures $X$ and $Y$ of a given microwave mode, as needed for qubit readout in circuit quantum electrodynamics~\cite{vijay_observation_2011}. Even then, the signal-to-noise ratio remains ultimately limited by vacuum fluctuations enforced by Heisenberg uncertainty relations, imposing that the quadrature standard deviations satisfy $\delta X = \delta Y = 1/2$. As a result, a linear amplifier is not well suited for detecting fluorescence signals consisting of a few incoherent photons emitted randomly over many modes. In contrast, such signals are ideally detected by a photon counter: because it measures in the energy eigenbasis, it is in principle noiseless when the field is in vacuum and only clicks for photons incoming within its detection bandwidth~\cite{lamoreaux_analysis_2013}. 

{Operational Single Microwave Photon Detectors (SMPDs) have been developed only recently based on cavity and circuit quantum electrodynamics~\cite{gleyzes2007quantum, chen2011microwave, inomata_single_2016,narla2016robust, besse_single-shot_2018, kono2018quantum, lescanne_irreversible_2020}. Here, we report the first use of such a SMPD for sensing applications, to detect spin fluorescence at microwave frequencies.} Consider an ensemble of $N$ electron spins 1/2, resonantly coupled with a spin-photon coupling constant $g_0$ to a resonator of frequency $\omega_0$ and linewidth $\kappa$  (Fig.~\ref{fig1}a). After being excited by a $\pi$ pulse, the spins will relax exponentially into their ground state with a characteristic time $T_1$. If  their dominant relaxation channel is radiative (the so-called Purcell regime~\cite{bienfait_controlling_2016}), they will do so by spontaneously emitting $N$ microwave photons at the Purcell rate $\Gamma_P = 4g_0^2 / \kappa = T_1^{-1}$. Whereas a linear amplifier can only detect these incoherent photons as a slight increase of noise above the background~\cite{sleator_nuclear-spin_1985,mccoy_nuclear_1989}, a single-photon counter with sufficient bandwidth is expected to detect each of them as a click occurring at a random time and revealing an individual spin-flip event.

For our demonstration, we use the electronic spins of an ensemble of bismuth donors implanted about \SI{100}{nm} below the surface of a silicon chip enriched in the nuclear-spin-free silicon 28 isotope.
These spins couple magnetically to the inductor of a superconducting LC resonator with frequency $\omega_0/2\pi = \SI{6.94}{GHz}$ patterned in aluminium on the surface of the chip (see Fig.~\ref{fig1}b). 
Applying a static magnetic field $B_0\approx \SI{17}{mT}$ parallel to the inductor tunes the lowest transition frequency of the bismuth donors in resonance with the resonator. At this field, the energy loss rate $\kappa_i$ in the resonator is 3.5 times higher than the energy leak rate $\kappa_c$ in the measuring line, yielding a total resonator bandwidth $\kappa/2\pi = \SI{0.68}{MHz}$ with $\kappa=\kappa_i+\kappa_c$. As the electron spin of donors in silicon have hour-long spin-lattice relaxation times at low temperatures~\cite{tyryshkin_electron_2012}, they easily reach the Purcell regime when coupled to micron-scale superconducting resonators [\onlinecite{bienfait_controlling_2016}]. For our sample parameters, we measure a spin relaxation time $T_1 = 300 \pm 10 \ \text{ms}$ (see Sec. 3 in Methods) dominated by the radiative contribution. In our experiment [see Fig.\ref{fig1}(a)], the resonator coupled to the spins has a single input-output port connected through a circulator (and coaxial cables) to both the line used to drive the spins and to a SMPD.

This SMPD (see Fig.~\ref{fig1}c) consists of a superconducting circuit with a transmon qubit~\cite{koch_charge-insensitive_2007} of frequency $\omega_q$ capacitively coupled to two coplanar waveguide resonators: a 'buffer' resonator whose frequency $\omega_\mathrm{b}$ can be tuned to $\omega_0$ by applying a magnetic flux to an embedded superconducting quantum interference device (SQUID)~\cite{palacios-laloy_tunable_2008}, and a 'waste' resonator with fixed frequency $\omega_\mathrm{w}$. As described in Ref.~\cite{lescanne_irreversible_2020}, the detection of a photon (here at frequency $\omega_\mathrm{b}$) relies on the irreversible excitation of the transmon when driven by a non-resonant pump tone at frequency $\mathrm{\omega_p = \omega_q + \omega_w - \omega_b}$. The SMPD is cycled continuously, each cycle consisting of three steps (see Fig.~\ref{fig1}c).
First, a reset step ($R$), during which the qubit is set to its ground state by turning on the pump (violet pulse) while applying to the waste resonator a weak resonant coherent tone (green pulse).
Second, a detection step ($D$) that starts when the microwave at $\omega_\mathrm{w}$ is switched off, while the pump is kept on: a photon possibly entering the buffer gets mixed with the pump through a four-wave mixing process that triggers both the excitation of the transmon and the creation of a photon in the waste; this photon is lost in the \SI{50}{\ohm} port of the waste, which guarantees the irreversibility of the detection and the mapping of the incoming photon into a transmon excitation.
The third step ($M$) is the measurement of the transmon state using the dispersive shift~\cite{mallet_single-shot_2009} of the buffer resonator (orange pulse). At $\omega_\mathrm{b} = \omega_0$, the probability to detect a click when one photon reaches the detector during the detection window $D$ (intrinsic photon detector efficiency) is measured to be $\eta_\mathrm{d}=0.53 \pm 0.1$ (mainly limited by the transmon energy relaxation, see Methods), and the rate of false positive detection, referred to as the the dark count rate, is $\alpha = \SI{1.53}{clicks/ms}$. The detection duty cycle (step $D$ duration over total cycle duration) is $\eta_\mathrm{duty} = 0.43$, with a complete cycle lasting \SI{11.7}{\micro\second}. Note that the detector evidently saturates for signals having more than 1 photon every $\SI{10}{\micro s}$, approximately. The detector bandwidth $\Delta \omega / 2 \pi \approx \SI{2}{MHz}$ is larger than the spin resonator linewidth $\kappa/2 \pi$, implying that there is no filtering of the photons emitted by the spins.

\begin{figure}
\includegraphics[width=\columnwidth]{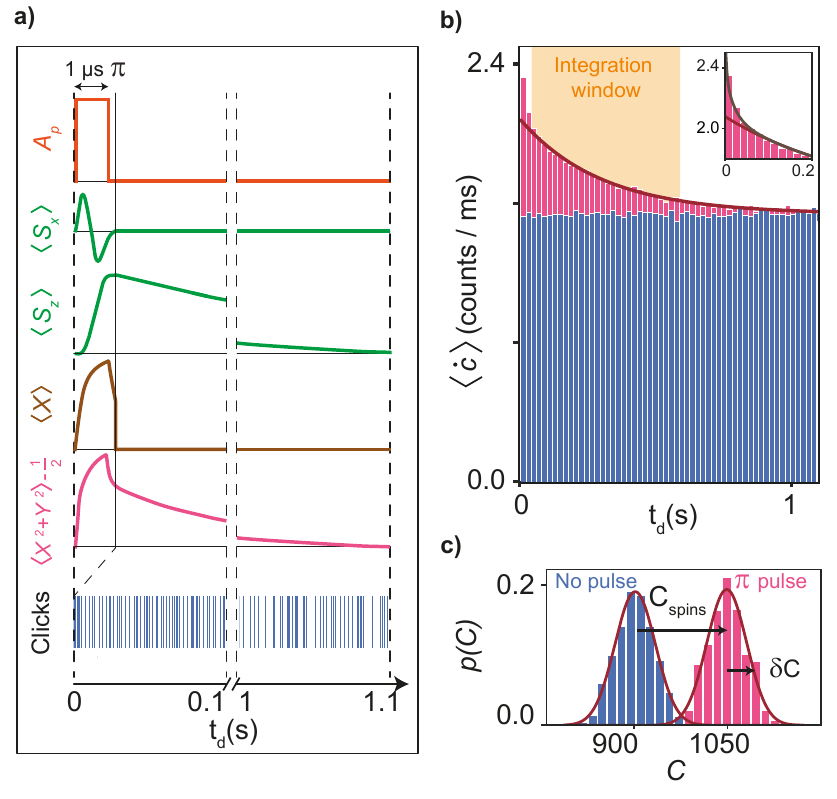}
 \caption{\textbf{Detection of spin relaxation by photon counting.}
 (\textbf{a}) Pulse sequence and spin dynamics. The applied pulse amplitude $A_p$, transverse and longitudinal magnetizations $\langle S_x \rangle$ and $\langle S_z \rangle$, output field quadrature $\langle X \rangle$, and photon count rate $\langle \dot{c} \rangle$ are shown as a function of the delay $t_d$ after the $\pi$ pulse. Solid curves are sketches of the expected dynamics. After a transient during the pulse, $\langle S_x \rangle$ quickly goes to $0$ because of the spin ensemble inhomogeneous broadening; $\langle S_z \rangle$ on the other hand is inverted and relaxes towards equilibrium in a characteristic time $T_1 = 0.3$\,s. During this time, $\langle X \rangle$ is also $0$, whereas a flux of spontaneous photons (the spin fluorescence) $\langle X^2 + Y^2 \rangle - 1/2$ is emitted. The lower panel shows one typical experimental time trace of the detected clicks as a function of $t_d$. 
 (\textbf{b}) Average count rate $\langle \dot{c} \rangle (t_d)$ measured in 19 ms time bins in the case where a $\pi$~pulse is (magenta) or is not (blue) applied to the spins. An exponential fit for $t_d>\SI{46.8}{ms} $ (solid line) leads to the characteristic time $T_1=\SI{309}{ms}$. The observed excess rate at short times {$t_d<\SI{50}{ms} $} (inset) has been investigated and attributed to fast-relaxing two-levels-systems.
 (\textbf{c}) Measured probability distribution of the number of counts $C$ integrated from \SIrange{46.8}{585}{ms}, obtained for 500 repetitions of the experiment when a $\pi$~pulse is either applied (magenta) or not (blue) to the spins. Solid lines represent Poissonian fits.
 }
\label{fig2}
\end{figure}

\begin{figure}
\includegraphics[width=\columnwidth]{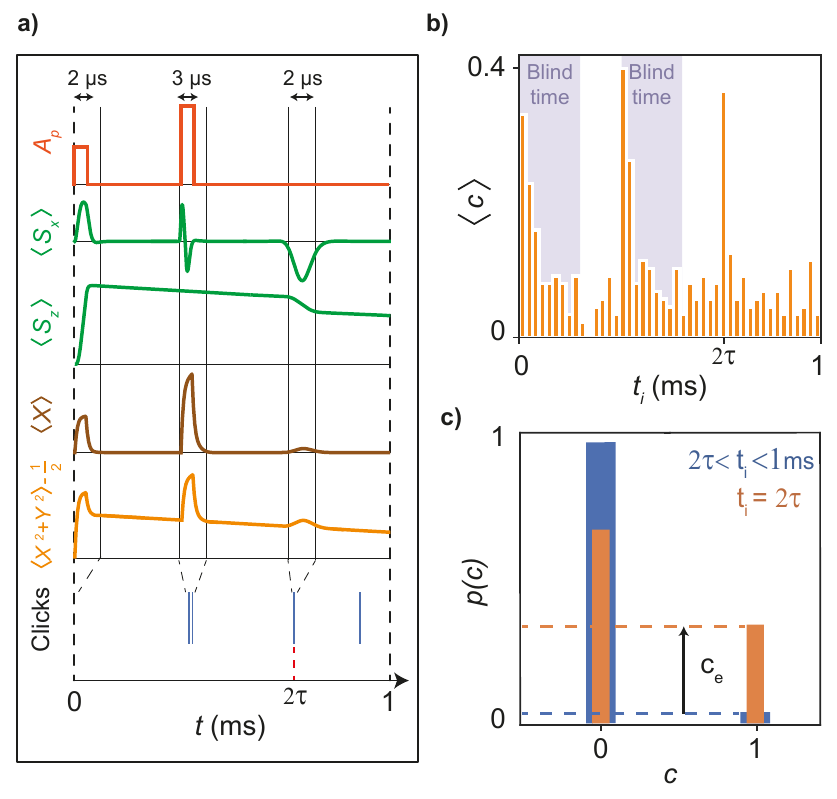}
 \caption{\textbf{Detection of spin echo by photon counting.}
 (\textbf{a}) Pulse sequence and spin dynamics. A Hahn echo pulse sequence (solid orange line) consisting of a $\pi/2$ pulse, a delay $\tau$, and a $\pi$ pulse is applied to the spin ensemble, causing a revival of the spin transverse magnetisation $\langle S_{x}\rangle$ at time $2\tau$, and the emission of a coherent microwave echo on the $X$ quadrature. After the $\pi/2$ pulse, the spin population $\langle S_z \rangle$ slowly decays by spontaneous emission with a characteristic time $T_1$, with in addition a small drop during the echo. Being coherent, the echo can be detected by homodyne detection or by a SMPD. Solid curves are sketches of the expected dynamics. An example of an experimental sequence is shown in the bottom, with a click detected at the echo time. 
 (\textbf{b}) Average number of counts $\langle c \rangle$ in $\SI{23}{\micro s}$ bins, as a function of the time $t$ from the beginning of the echo sequence, averaged over 83 sequences. Blue-shadowed areas represent the \SI{200}{\micro s} blind time of the detector after each strong spin pulse. The increased count probability at $t_i=2\tau$ is the spin-echo.
 (\textbf{c}) Average probability $p(c)$ of having one or no count in the time bin centred at echo time $t_i=2\tau$ (orange) or in one of the subsequent bins $2\tau<t_i<1 \, \text{ms}$ (blue). The difference between click probabilities (dashed lines) lead to a signal $c_{e}$ of 0.3 photons re-emitted coherently.
 }
\label{fig3}
\end{figure}

As a first experiment, we measure the spontaneous emission of the spin ensemble: a $\pi$-pulse  inverts the spin population, so that $\langle S_z \rangle = N/2$, and $\langle S_x \rangle =  \langle S_y \rangle = 0$, $S_{x,y,z} = \sum_{i=1}^N S^{(i)}_{x,y,z}$ being the sum of the individual dimensionless spin operators $\mathbf{S}^{(i)}$ (Fig.~\ref{fig2}a). Since $\langle S_{x,y} \rangle = 0$, the coherent part of the output field also satisfies $\langle X \rangle  = \langle Y \rangle = 0$. On the other hand, by energy conservation, a flux of incoherent photons is emitted at a rate $\langle X^2 + Y^2\rangle - 1/2$ proportional to $- \partial_t \langle S_z \rangle$ (see Fig.~\ref{fig2}a), forming the spin fluorescence signal and triggering counts in the SMPD. This signal decays back to $0$ within the spin relaxation time $T_1$. Note that the $\pi$ pulse perturbs the SMPD during a dead-time of $\sim \SI{200}{\micro s}$, after which it can be used normally.

One measurement record consists of \num{2e5} consecutive SMPD detection cycles, spanning a total measurement time of \SI{2}{s}. Each cycle yields one binary outcome $c(t_i)$, $t_i$ being the time around which cycle $i$ is centered. Figure ~\ref{fig2}a displays an example of a measurement record at early (\SIrange{0}{0.1}{s}) and at late times (\SIrange{1}{1.1}{s}): due to the emission by the spins, more counts are observed in the \SIrange{0}{0.1}{s} interval than in \SIrange{1}{1.1}{s}. 
Repeating the measurement 500 times and {histogramming the number of counts}, we obtain the average count rate $\langle \dot{c} (t_d)\rangle$ as a function of the delay $t_d$ after the $\pi$ pulse. Figure \ref{fig2}b shows this rate with and without $\pi$-pulse applied. Without pulse, a constant $\langle \dot{c}  (t_d) \rangle= \SI{1.53}{counts/ms}$ is recorded, which corresponds to the dark count rate $\alpha$ of the SMPD. With $\pi$-pulse, $\langle \dot{c}  (t_d)\rangle$ shows an excess of \SI{0.85}{counts/ms} exponentially decaying towards $0$, with a fitted time constant of \SI{309}{ms}. Because this time is the same as the independently measured spin relaxation time $T_1$ (see Methods), we conclude that the SMPD detects the photons spontaneously emitted by the spins upon relaxation. Note that at short $t_d$ we moreover observe an extra excess rate of \SI{0.3}{counts/ms} decaying with a \SI{20}{ms} time constant (see inset of Fig.~\ref{fig2}b), which we attribute to the radiative relaxation of spurious two-level systems present at sample interfaces.

To analyse the photo-counting statistics of the fluorescence signal, we integrate the number of counts $C = \sum_i c(t_i)$ {over a window of duration $t_w = 540$\,ms}. The probability histogram $p(C)$ is shown in Fig.~\ref{fig2}c. With and without $\pi$ pulse, an average of $\langle C  (\pi)\rangle = 1050$ and $\langle C (0)\rangle = 900$ counts are detected, the difference defining the spin signal $C_\mathrm{spin} = 150$ photons. The ratio of this signal to the total number of excited spins $N$ defines an overall detection efficiency $\eta =C_\mathrm{spin}/N$. For a Poissonian distribution, one expects the width of $p(C)$ to be $ \delta C = \sqrt{\alpha t_w}$ and $\delta C = \sqrt{\alpha t_w + \eta (1-\eta) N }$ without and with $\pi$ pulse, respectively. In our case, because $\alpha t_w \gg \eta N$, both distributions have approximately the same width $\delta C \simeq 30$ counts, dominated by the dark counts fluctuations contribution. 

It is interesting to note that the signal-to-noise ratio $\eta N / \sqrt{\alpha t_w + \eta (1-\eta) N}$ can {in principle} become arbitrarily large for an ideal SPD for which $\alpha \sim 0$ and $\eta \sim 1$, even for $N$ approaching $1$. This reflects the fact that in the Purcell regime, $N$ spins once excited will emit $N$ photons over a timescale of a few $T_1$, and that an ideal SMPD will detect them all noiselessly. This is in marked difference with previous experiments using superconducting qubits for ESR spectroscopy, which relied on either Free-Induction-Decay collection by a tunable resonator~\cite{kubo_electron_2012} or measurement of the spin ensemble magnetisation by a flux-qubit~\cite{toida_electron_2019,budoyo_electron_2020}. SMPD detection of spin fluorescence in the Purcell regime thus appears as a particularly promising method for detecting small numbers of spins. In our experiment, the SNR is equal to $4.6$, already exceeding the SNR of echo-based detection as discussed in the following, despite the imperfections of the present SMPD.

Using additional spin measurements by homodyne detection~\cite{schweiger_principles_2001} combined with a numerical simulation of the experiment, we estimate that $N = (13.3 \pm 1.1) \times 10^3$ spins are excited by the $\pi$ pulse (Sup. Mat.). The overall detection efficiency is thus $\eta =  0.011 \pm 0.003$. Writing this efficiency as $\eta = \eta_\mathrm{d} \eta_\mathrm{duty} \eta_\mathrm{int} \eta_\mathrm{col}$, with $\eta_\mathrm{int}=0.71$ a factor due to the finite integration window, we deduce a collection efficiency $\eta_\mathrm{col}= 0.07 \pm 0.015$ between the spins and the detector. This is due in part to the spin resonator internal losses which contribute for a factor $ \kappa_\mathrm{c}/(\kappa_\mathrm{i} + \kappa_\mathrm{c}) = 0.22$, and in part to losses in the microwave circuitry joining the two devices.

\begin{figure}
\includegraphics[width=\columnwidth]{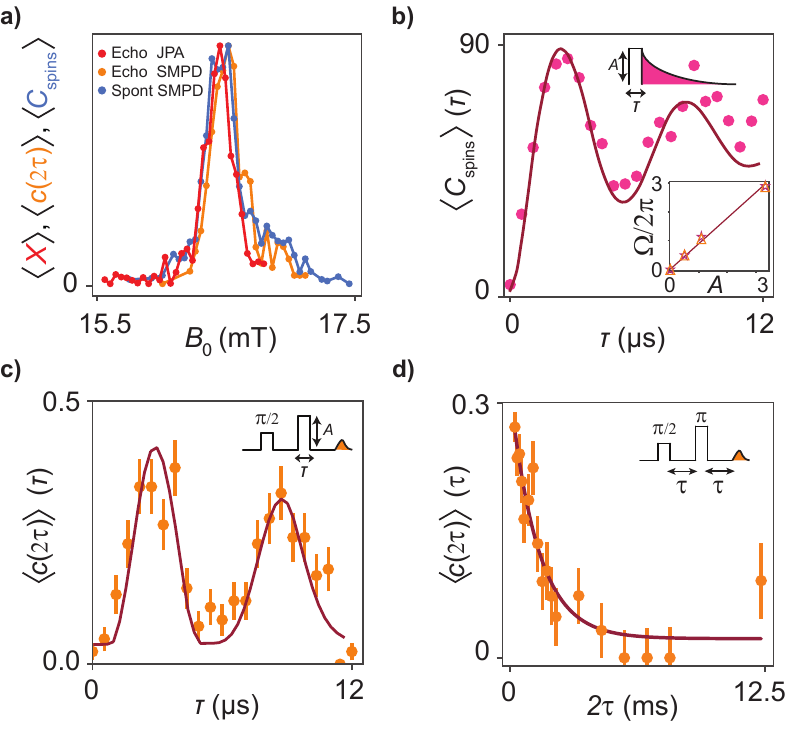}
 \caption{\textbf{Spin ensemble characterisation by photon counting.}  
 (\textbf{a}) Spin ensemble resonance lineshape when scanning the magnetic field $B_0$ measured with the three spin detection methods: homodyne echo detection (red curve), echo detection by photon counting (orange curve) and spontaneous emission detection by photon counting (blue curve).
 (\textbf{b}) Measured (magenta dots) and simulated (solid line) average spin signal $\langle C_\mathrm{spin}\rangle$ as a function of the duration $T$ of a microwave pulse exciting the spins.
 (\textbf{c}) Measured (orange dots) and simulated (solid line) average number of clicks $\langle c(2\tau)\rangle$ detected at echo time, as a function of the duration $T$ of the second pulse of the Hahn echo sequence. The extracted Rabi frequency dependence on the pulse amplitude $A$ (inset of panel b) matches the one obtained through the spontaneous emission signal.
 (\textbf{d})Measured (orange dots) average number of photons detected at echo time $\langle c(2\tau)\rangle$ as a function of the time delay $\tau$ between the pulses of the Hahn echo sequence.
 An exponential fit (solid line) yields a coherence time $T_2 = \SI{2.7}{ms}$. 
 }
\label{fig4}
\end{figure}

We now turn to another method of spin detection by a SMPD, during the emission of a spin echo at the end of a Hahn echo sequence $\pi/2 - \tau - \pi - \tau - echo$ (see Fig. \ref{fig3}a). After the first $\pi/2$~pulse, which brings all the spins along the $x$ axis at $t=0$, spins lose phase coherence in a time $\sim T_E$ due to the spread of their Larmor frequencies, so that $\langle S_x  (t) \rangle \sim (N / 2) \mathrm{e} ^{-2t/T_E}$. In our experiment, $T_E \sim \kappa^{-1}$ because the spin excitation bandwidth is set by the cavity and not by the much larger spin ensemble inhomogeneous linewidth (Fig.~\ref{fig4}). Phase coherence is transiently restored around $t = 2\tau$ by the refocusing $\pi$ pulse, yielding $\langle S_x  (t) \rangle \sim (N/2) \mathrm{e} ^{-2|t-2\tau|/T_E}$. The oscillating transverse magnetization generates a short phase-coherent microwave pulse of duration $T_E$ in the detection line, called the spin echo, with the photon statistics of a coherent state. In the limit $N \Gamma_P T_E \ll 1$, its amplitude can be shown to be $\langle X_e \rangle \sim  N \sqrt{{\eta_\mathrm{col} \Gamma_P T_E/2}}$, corresponding to an average photon number $\langle X_e \rangle^2$~\cite{bienfait_reaching_2016} much smaller than the number $N$ of spins. Spin echoes are usually detected by linear amplification and phase-coherent demodulation~\cite{schweiger_principles_2001,bienfait_reaching_2016}, with a signal-to-noise ratio  $\langle X_e \rangle / \delta X = 2 \langle X_e \rangle$ ultimately limited by the vacuum fluctuations~\cite{bienfait_reaching_2016}.
Here, we show that spin echoes can also be detected by a microwave SMPD, as photon echoes at optical frequencies~\cite{abella_photon_1966}. Note that the signal-to-noise ratio upper-bound that an ideal SMPD could reach is limited by photon shot noise during the echo and equal to $\langle X_e \rangle^2 / \sqrt{\langle X_e \rangle^2} = \langle X_e \rangle$, i.e. half the one of phase-coherent detection.

In our demonstration of microwave photon echo detection, the echo duration is shorter than the detection cycle, and only one photon at most can be detected at the echo time $2\tau${; we thus} center the detection step $D$ of the SMPD at $2\tau$. We also chose $\tau = \SI{350}{\micro s}$, larger than the detector dead time. A typical photo-counting trace is visible in Fig.~\ref{fig3}a, showing in particular one click at the expected echo time. Repeating several echo sequences yields $\langle c(t_i) \rangle$ (see Fig.~\ref{fig3}b), clearly showing an excess of counts for $t_i = 2 \tau$.

The click probability histogram is shown in Fig.~\ref{fig3} at and out of the echo time. The average number of detected photons during the spin-echo, $c_\mathrm{echo} = \langle c(2\tau) \rangle - \langle c(t_i>2\tau) \rangle = 0.3$, is as expected much lower than $C_\mathrm{spin}$, the number of photons detected in the spontaneous emission experiment of Fig.~\ref{fig2}.
The standard deviation $\delta c_\mathrm{echo} = 0.46$ during the echo (Fig.~\ref{fig3}) yields a signal-to-noise ratio $c_\mathrm{echo}/\delta c_\mathrm{echo} = 0.65$, significantly lower than the one obtained with the spontaneous emission method, although both measurements were performed with the same repetition time $\sim 2 T_1$ and thus also the same initial spin polarization.

We finally demonstrate that SMPD detection can be used to perform usual spin characterisation measurements. First, spin spectroscopy is performed by varying the magnetic field $B_0$ around the resonance value and using the three different detection methods already mentioned: homodyne detection with echo, SMPD {fluorescence detection}, and SMPD detection of the echo. As seen in Fig.~\ref{fig4}d, all three methods give similar spectra.
Second, we observe Rabi nutations in the {fluorescence} signal. In Fig.~\ref{fig4}b, the spin signal $C_\mathrm{spin}$ is plotted as a function of the spin driving pulse duration $\tau$. Oscillations are observed, with a frequency linearly dependent on the pulse amplitude, reflecting the Rabi oscillations of $\langle S_z \rangle$ since $C_\mathrm{spin} = \eta [1 + 2 \langle S_z \rangle]/2$. This Rabi nutations can also be measured with the microwave photon echoes method, by varying the duration $\tau$ of the refocusing pulse; the SMPD signal $\langle c  (2\tau) \rangle$ shows the expected Rabi oscillations (Fig.~\ref{fig4}c). The oscillation contrast in Figs.~\ref{fig4}b and c diminishes with pulse duration $\tau$ due to the spread of Rabi frequencies in the ensemble. This inhomogeneity has a different impact on the spontaneous emission signal and on the echo signal~\cite{ranjan_pulsed_2020}, which is quantitatively reproduced by simulations (see Methods), as seen in Figs~\ref{fig4}b-c. Finally, the spin coherence time is measured by microwave photon echo detection. In Fig.~\ref{fig4}d, $\langle c  (2\tau) \rangle$ is plotted as a function of $\tau$. An exponential fit to the data yields $T_2 = 2.7 $\,ms, in agreement with the value measured using homodyne detection (Sup. Mat.). Overall, this demonstrates that SMPD detection can be used to perform standard ESR spectroscopy measurements.

Our results represent the first use of a SMPD for quantum sensing. Beyond the fundamental interest of such a proof of principle, we conclude by discussing the potential of SMPDs {for spin detection}. Whereas the SNR of echo detection by a SMPD (Fig.~\ref{fig3}) and by homodyne detection are comparable, spin fluorescence detection by a SMPD (Fig.~\ref{fig2}) on the other hand presents several features that make it a truly interesting method for ESR spectroscopy. Indeed, the SNR can reach much higher values than in echo-detection, since the signal (number of emitted photons) can be as high as the total number of excited spins, whereas the noise is entirely dominated by SMPD non-idealities, which are likely to be improved in future devices~\cite{royer_itinerant_2018-1,grimsmo_quantum_2020,kokkoniemi_bolometer_2020}. We therefore expect that the development of better SMPDs with lower dark count rates and higher efficiency will push further ultra-sensitive spin detection, possibly down to a single spin. This perspective is all the more interesting that the method applies equally well to spins with short coherence times such as encountered in real-world spin systems, making practical single-spin ESR spectroscopy a possible future perspective.

\subsection*{Acknowledgements}
{We acknowledge technical support from P.~S\'enat, D. Duet, P.-F.~Orfila and S.~Delprat, and are grateful for fruitful discussions within the Quantronics group. This project has received funding from the European Unions Horizon 2020 research and innovation program under Marie Sklodowska-Curie Grant Agreement No. 765267 (QuSCO). E.F. acknowledges support from the ANR grant DARKWADOR:ANR-19-CE47-0004. We acknowledge support from the Agence Nationale de la Recherche (ANR) through the Chaire Industrielle NASNIQ  under contract ANR-17-CHIN-0001 cofunded by Atos, and of the Région Ile-de-France through the DIM SIRTEQ (REIMIC project).}

\subsection*{Author contributions}

E.A., P.B. and E.F. designed the experiment. T.S. provided the bismuth-implanted isotopically purified
silicon sample, on which V.R. fabricated the Al
resonator. E.A. designed and fabricated the SMPD with the help of D.V. and E.F.. E.A., V.R., E.F. performed the measurements, with help from L.B, D.F. and P.B.. E.A., P.B. and E.F. analysed the data. E.A., E.B. and V.R. performed the simulations. E.A., P.B. and E.F. wrote the manuscript.
D.F., D.V., D.E. and E.F. contributed useful input to the manuscript.

\bibliographystyle{ieeetr}
\bibliography{SpinSPD}

\clearpage

\section{\Large{METHODS}}

\section{1. Setup}
The experimental setup used to drive the spin ensemble and operate the SMPD, with six input lines (labeled 1-6) and one output line (labeled 7), is shown in Fig.~\ref{Setup}. We first discuss the room-temperature part (microwave and dc signals generation), and then the low-temperature part (cabling inside the dilution refrigerator).

\subsection{Room temperature setup}

The room-temperature setup includes four microwave sources and two 4-channel arbitrary waveform generators (AWG 5014 from Tektronix). All microwave pulses needed in the experiment are generated by mixing the output of a source with $2$ AWG channel outputs used to drive the I and Q ports of an I/Q mixer at an intermediate frequency indicated in Fig.~\ref{Setup}. The pulses are used to drive the spins (at the spin resonator frequency $\omega_0$) and to operate the SMPD. 

SMPD operation requires:

- a dc flux-bias of the SQUID in the buffer resonator, in order to tune $\omega_b$ in resonance with $\omega_0$. This is achieved with a dc current source  (Yokogawa 7651) connected to an on-chip antenna near the SQUID (line 4 in Fig.~\ref{Setup}).

- microwave pulses at the pump frequency $\omega_p$ to satisfy the 4-wave mixing condition $\mathrm{\omega_p=\omega_q+\omega_w-\omega_b}$

- microwave pulses to readout the qubit state via the qubit-state-dependent dispersive shift of the buffer resonator. They are at frequency $\omega_b + \chi_{qb}$, the buffer resonator frequency with qubit in the $e$ state. 

- microwave pulses at the waste frequency $\omega_w$ to reset the qubit.

Moreover, qubit readout pulses are amplified by a flux-pumped Josephson Parametric Amplifier (JPA) in degenerate mode~\cite{zhou_high-gain_2014}. The JPA needs dc flux biasing to adjust the JPA frequency; it is provided by an on-chip antenna near the JPA SQUID array, fed by a constant voltage source biasing a resistor at room-temperature (line 1 in Fig.~\ref{Setup}). The JPA also requires flux-pumping to achieve gain. The pump tone is generated by frequency-doubling the same source used to generate the readout pulses (line 2 in Fig.~\ref{Setup}), followed by mixing with an intermediate frequency (see Fig.~\ref{Setup}). The relative phase between signal and pump is adjusted with a phase shifter for maximum gain on the signal-bearing quadrature. 

The same source (Keysight MWG, shown in yellow in Fig.~\ref{Setup}) is used for driving the spins, qubit state readout, JPA pumping, and as local oscillator for signal demodulation yielding the quadratures of the qubit readout pulses. Spin driving pulses and qubit state readout pulses are sent via the same line (line 3 in Fig.~\ref{Setup}). Spin driving pulses require much larger powers than qubit readout pulses. Therefore, in the room-temperature setup, the line was split before recombination, and in one of the branches an amplifier was inserted in-between two microwave switches. 

A second source (Vaunix Labbrick, shown in green in Fig.~\ref{Setup}) is used for the qubit reset pulses at $\omega_w$ (line 5). A third source (Keysight, shown in purple in Fig.~\ref{Setup}) is used for SMPD pumping. Pump pulses are generated through I/Q mixing and amplification of the generator output. The signal is then passed through a \SI{70}{MHz} band-pass filter to prevent spurious wave mixing caused by side-band resonances and LO leakage, before reaching the cryostat input on line 6. A fourth source (Vaunix Labbrick, shown in blue in Fig.~\ref{Setup}) is used for SMPD tuning and characterization.

\subsection{Low-temperature setup}

\begin{figure*}
\includegraphics[width=0.9\textwidth]{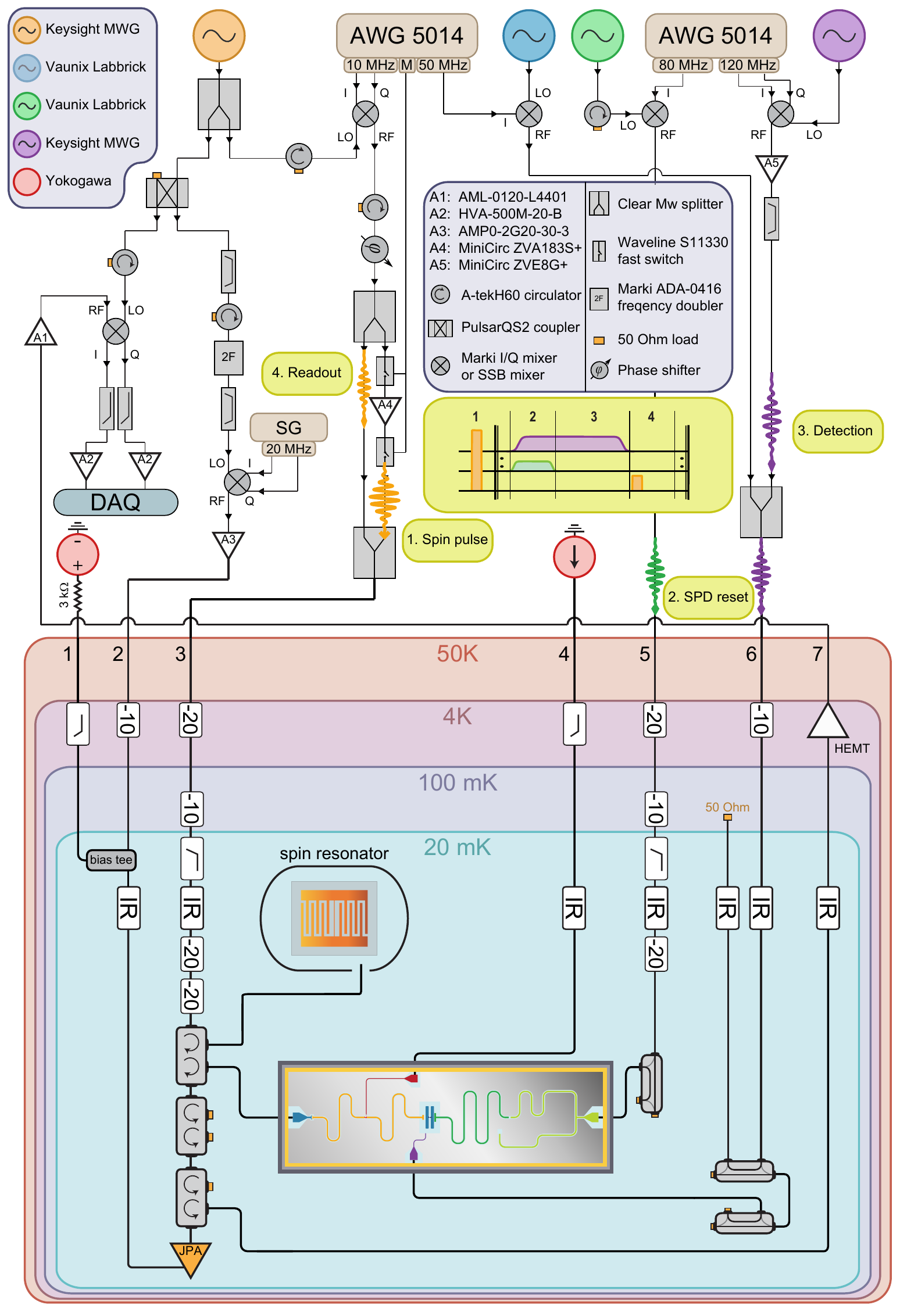}
\caption{\textbf{Schematic of the setup.} }
\label{Setup}
\end{figure*}

Line 3 is heavily attenuated at low-temperatures in order to minimize spurious excitations of the transmon qubit in the SMPD and therefore dark counts (see Fig.~\ref{Setup}). It is then connected to the spin resonator input via a double circulator. The reflected signal is routed by the same circulator towards the SMPD input (buffer resonator), and the signal reflected on the SMPD is finally routed towards the input of the JPA and the detection chain. Two double circulators isolate the SMPD from the JPA, to minimize noise reaching the SMPD and potentially causing spurious qubit excitations and dark counts. The JPA output (reflected signal) is routed to a High-Electron-Mobility-Transistor (HEMT) amplifier from Low-Noise Factory anchored at the 4K stage of the cryostat, and then to output line 7. Infrared filters are inserted on all the lines leading to the SMPD to minimize out-of-equilibrium quasi-particle generation leading to spurious qubit excitations and dark counts. To minimize heating of the low-temperature stage by the strong pump tone of the SMPD, the necessary attenuation of the pump line at $10$\,mK is achieved with a $20$\,dB directional coupler that routes most of the pump power towards the $100$\,mK stage where it is dissipated.

Using the same line both for spin excitation and SMPD readout raises potential issues that are now discussed. First, the spin excitation pulse also leads to a large field build-up in the buffer resonator (since $\omega_b = \omega_0$), which excites the qubit and perturbs the proper functioning of the SMPD during a time that we quantify to be $200 \mu \mathrm{s}$ (detector dead-time). Then, one may also wonder about spurious excitation of the spins caused by the repeated qubit readout pulses. This is avoided, because qubit readout is performed at $\omega_b + \chi_{qb}$, which is thus shifted from $\omega_0$ by $\chi_{qb}/2\pi = -3.5$\,MHz.

\section{2. Fabrication}

\subsection{Spin sample} \label{spin_sample}
The bismuth donors are implanted in a $700$\,nm epilayer of $^{28}\mathrm{Si}$-enriched silicon. The implantation profile ranges from 50 to 150nm depth, with a peak concentration of \SI{8e16}{donors/cm^3} (see refs~\cite{bienfait_reaching_2016,ranjan_electron_2020} for more details). The spin resonator consists of an interdigitated capacitor, shunted by a $1 \mu \mathrm{m}$-wide, $450 \mu \mathrm{m}$-long inductive wire, with a design similar to the one used in~\cite{bienfait_reaching_2016}. It is deposited on top of the silicon sample by evaporation of a \SI{50}{nm}-thick aluminium film through a resist mask patterned using e-beam lithography, followed by liftoff. 
The chip is then placed inside a 3D copper cavity into which a pin protrudes, controlling the capacitive coupling to the input line~\cite{bienfait_reaching_2016}.
A superconducting coil applies an in-plane magnetic field $\mathrm{B_0}$ to tune the spin frequency.

\subsection{Single microwave photon detector}
The single photon detector circuit is based on the design by Lescannne et al.~\cite{lescanne_irreversible_2020}. It is fabricated using wet etching of a \SI{60}{nm} aluminium layer evaporated on a high-resistivity intrinsic silicon substrate.
Before metal deposition, the substrate is pre-cleaned with a SC1 process.
The wafer is first immersed for \SI{10}{min} at \SI{80}{\degree C} in a bath of
5 parts $\mathrm{H}_2 \mathrm{O}$ to 1 part $\mathrm{H}_2 \mathrm{O}_2 \, (\SI{30}{\percent})$ to 1 part $\mathrm{N} \mathrm{H}_4 \mathrm{OH} \, (\SI{29}{\percent})$, then is immersed for \SI{2}{min} in HF (\SI{5}{\percent}) solution to remove the surface oxide.
The substrate is then loaded in an electron-beam evaporator within \SI{10}{minutes}, after which a \SI{60}{nm} aluminium layer is deposited.
Patterning of the circuit is achieved by electron beam lithography of a UV3 resist mask, followed by wet etching of the aluminium using a TMAH-based developer (Microposit CD26).
The Josephson junctions are evaporated using the Dolan bridge technique and recontacted to the main circuit through aluminium bandage patches~\cite{dunsworth_characterization_2017}.
Finally, circuit gaps are isotropically trenched with a SF6-based reactive ion etch, which has shown to decrease the internal losses of superconducting resonators~\cite{bruno_reducing_2015,calusine_analysis_2018}. The resulting \SI{10}{mm} by \SI{3}{mm} chip is glued and wired to a Printed-Circuit-Board, placed in a copper box, magnetically shielded, and attached to the cold stage of the dilution refrigerator.

\section{3. Characterisation}

\subsection{Electron spin resonance spectroscopy by homodyne measurements}

Prior to the experiments reported in the main text, the spin ensemble is characterised by pulsed electron spin resonance spectroscopy, comparable to previous work~\cite{bienfait_reaching_2016}. Interestingly this can be done in the same cooldown as the SMPD measurements reported in the main text, because of the fact that the SMPD setup also includes a Josephson Parametric Amplifier (JPA) for qubit state readout. 
To switch from single photon detection to homodyne spin measurements, we simply tune the buffer resonator frequency $\omega_b$ at a frequency far from $\omega_0$, and tune the JPA at resonance with $\omega_0$. In that way, the spin-echo signal simply reflects off the SMPD without triggering any qubit excitation, and gets amplified by the JPA, exactly as was achieved in similar experiments \cite{bienfait_reaching_2016}. Output signal demodulation then yields the spin-echo quadrature and its integral $A_e$.

\begin{figure}[h]
\includegraphics[width=\columnwidth]{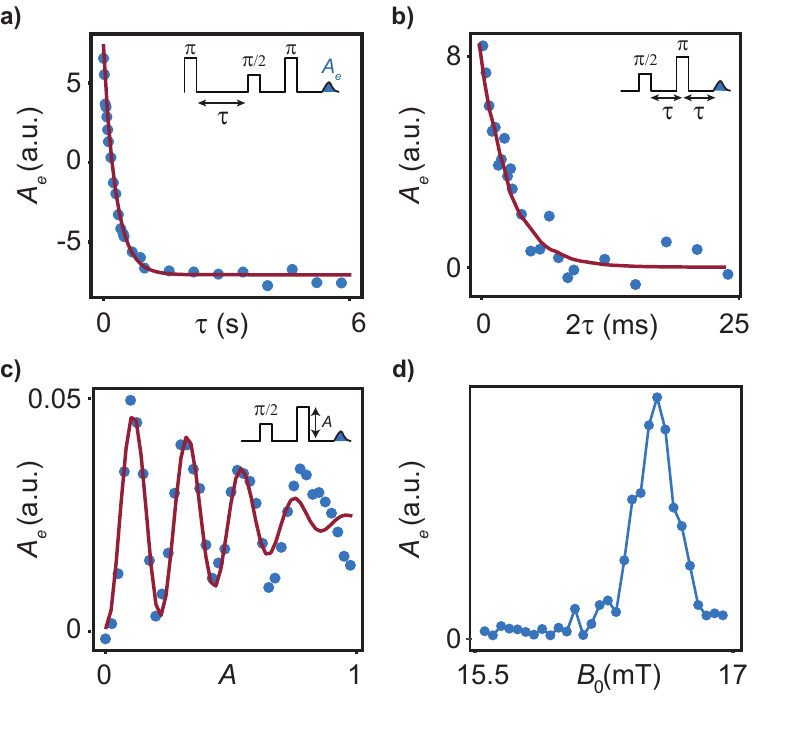}
 \caption{\textbf{ESR spectroscopy of the spin ensemble.}  
 (\textbf{a}) Measured (blue dots) and simulated (solid line) integrated echo  as a function of the delay $\tau$ between the inversion $\pi$-pulse and the Hahn echo sequence. An exponential fit (not shown) yield a characteristic decay time $T_1 = 300 \pm 10 \, \text{ms}$.
(\textbf{b}) Measured (blue dots) and simulated (solid line) integrated echo  as a function of the delay $\tau$ between $\pi/2$ and $\pi$ pulses of the Hahn echo sequence. An exponential fit (not showed) yields a characteristic decay time $T_2 = \SI{2.7}{ms}$.
 (\textbf{c})  Measured (blue dots) and simulated (solid line) integrated echo  as a function of the amplitude $A$ of the $\pi$ pulse of the Hahn echo sequence revealing Rabi oscillations. 
 (\textbf{d}) Measured integrated echo (blue dots) as a function of the in-plane magnetic field $\mathrm{B_0}$ used to tune the spin ensemble frequency.
 }
\label{SupHahn}
\end{figure}

We measure the spin relaxation time $T_1$ at $B_0 = 17$\,mT with an inversion recovery sequence, in which a $\pi$ pulse is first applied, followed after a duration $\tau$ by a Hahn-echo detection sequence. The echo area $A_e$ is shown as a function of $\tau$ in Fig.~\ref{SupHahn}a, together with an exponential fit yielding $T_1 = 300 \pm 10 \, \text{ms}$. We also measure the spin coherence time by measuring the echo amplitude as a function of the delay $2\tau$  between the $\pi/2$ pulse and the echo (see Fig.~\ref{SupHahn}b). An exponential fit yields $T_2 = 2.7$\,ms. Rabi nutations are obtained by measuring the echo area $A_e$ as a function of the refocusing pulse amplitude $A$ (see Fig.~\ref{SupHahn}c). Finally, the bismuth donor spin spectrum is obtained by recording the echo amplitude $A_e$ as a function of the field $B_0$ (see Fig.~\ref{SupHahn}d).

The data in Fig.~\ref{SupHahn} are modelled using a simulation tool described elsewhere~\cite{ranjan_pulsed_2020}. It computes the evolution of the spin ensemble under the application of driving pulses at the resonator input. The spread in spin Larmor frequency (due to strain-induced inhomogeneous broadening~\cite{pla_strain-induced_2018}) and in spin-photon coupling (due to the spatial inhomogeneity of the $B_1$ field generated by the resonator) are taken into account by describing the spin as an ensemble of packets, with coupling constant density $\rho(g_0)$ and frequency density $\rho_\mathrm{spin}(\omega)$. The evolution of each packet is computed independently under the drive pulses, and the echo response is obtained by summing the packet contributions. Purcell relaxation is also taken into account~\cite{ranjan_pulsed_2020}. 

Here, we make two extra simplifying assumption. Because the inhomogeneous broadening is much larger than the cavity linewidth $\kappa$, and that the signal originates essentially from spins within this linewidth, we consider the spin density to be constant, $\rho_\mathrm{spin}(\omega) = \rho_\mathrm{spin}$. Moreover, we model the coupling constant inhomogeneity as a Gaussian centered on $\bar{g}_0$ and width $\delta g_0$, $\rho(g_0) =\frac{1}{\delta g_0 \sqrt{2\pi}}  e^{-\frac{(g_0 - \bar{g}_0)^2}{2\delta g_0 ^2}}$. We adjust the values of $\bar{g}_0$ and $\delta g_0$ to get good agreement with the relaxation and Rabi nutation data in Fig.~\ref{SupHahn}b and d, yielding $\bar{g}_0/2\pi = 290$\,Hz, and $\delta g_0 /2\pi = 25$\,Hz. Because the spin density $\rho_\mathrm{spin}$ only rescales the signal amplitude in Fig.~\ref{SupHahn}, its determination requires other measurements that are described below, enabling us to infer the number of excited spins and the overall photon detection efficiency. 

\subsection{Single photon detector characterisation and tuning}
The single microwave photon detector consists of a transmon qubit whose ground {$g$} and first excited state {$e$} encode the detector click. The transmon frequency
is $\mathrm{\omega_q} / 2 \pi = \SI{6.14}{GHz}$
and its anharmonicity is
$\SI{-200}{MHz}$. It is capacitively coupled to a tunable buffer resonator (maximum frequency $\omega_\mathrm{b}^\mathrm{max} / 2 \pi = \SI{7.09}{GHz}$)
and a waste resonator ($\omega_\mathrm{w} / 2 \pi = \SI{7.62}{GHz}$)
with dispersive shifts
$\chi_\mathrm{qb}/ 2 \pi = \SI{-3.5}{MHz}$
and
$\chi_\mathrm{qw} /2\pi= \SI{-8.1}{MHz}$ respectively.
The buffer resonator is coupled to the external microwave line via a capacitance
(energy damping rate $\kappa_\mathrm{b} = 13.7 \times 10^6\ \mathrm{s^{-1}}$ at the working point $\omega_b/2\pi=\SI{6.946}{GHz}$),
while the waste resonator is coupled through a Purcell filter 
to a \SI{50}{\ohm}-terminated line
(energy damping rate $\kappa_\mathrm{w} = 2.8 \times 10^6\ \mathrm{s^{-1}}$).

\begin{figure}
\includegraphics[width=\columnwidth]{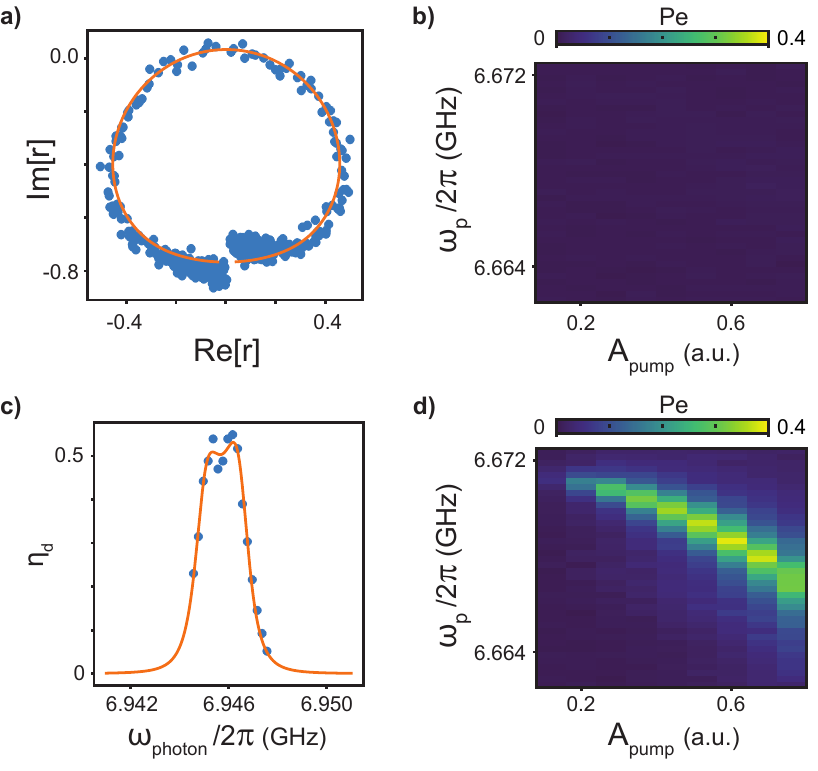}
 \caption{\textbf{Single microwave photon detector characterisation.}  
 (\textbf{a}) Measured (blue dots) and fitted (orange solid line) complex reflection coefficient $r$ of the buffer resonator at working point $\omega_b/2\pi=\SI{6.946}{GHz}$. The fitting function takes into account flux noise of the SQUID enabling the tuning the resonator.
 (\textbf{b}), (\textbf{d}) Probability $\mathrm{P_e}$ of finding the qubit in its excited state (color scale) as function of the amplitude $A_\mathrm{pump}$ and frequency $\omega_\mathrm{pump}/2\pi$ of the pump activating the parametric process of photo-detection.
 When no photon is impinging, the buffer resonator is close to its vacuum state (b) no parametric process is activated and the qubit is mostly in its ground state $\mathrm{P_e}\approx0$; in contrast, when photons are injected (d) the parametric process is activated at pump frequencies for which the conservation of energy is respected.
 The quadratic dependence of the pump activation frequency on the pump amplitude is due to the Stark shift of the qubit frequency for increasing pump power.
 (\textbf{c}) Measured (blue dots) efficiency of detection $\eta_\mathrm{d}$ at $\omega_\mathrm{b}=\SI{6.946}{GHz}$, as a function of the input photon frequency $\omega_\mathrm{photon}/2\pi$. From the fit (orange solid line), obtained with a model of two coupled cavities, we extract a bandwidth  $\Delta_\mathrm{det}/2\pi\approx \SI{2.1}{MHz}$. 
 }
\label{SupSMPD1}
\end{figure}

\begin{figure}
\includegraphics[width=\columnwidth]{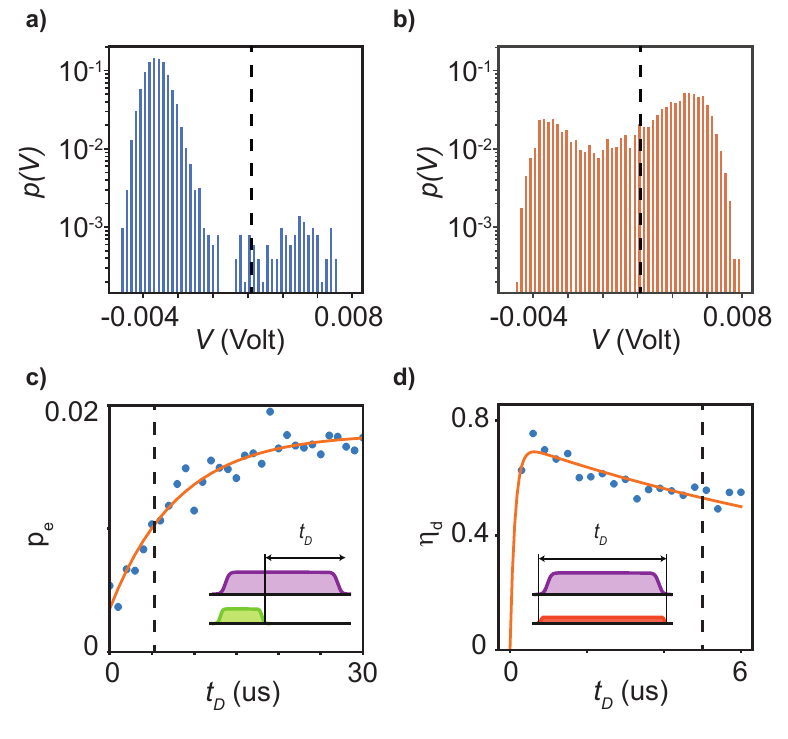}
 \caption{\textbf{SMPD performance}  
 Probability $p(V)$ of measuring the average quadrature voltage $V$ when probing the buffer resonator for qubit readout, when a pulse is applied (\textbf{b}) or not (\textbf{a}) to the qubit prior to the measurement. Dashed line indicates the readout threshold, chosen to minimise the ratio $\alpha/\eta_d$, each measure falling on the left (resp. right) is associated to the qubit being in its ground (resp. excited) state.
 (\textbf{c}) Measured (blue dots) and fitted (orange solid line) probability $p_e$ of finding the qubit in its excited state as function of time $T$ after the reset sequence, showing out-of-equilibrium qubit excited population reaching thermal equilibrium on a timescale $\approx T_1$. Black dashed line at $t_D=\SI{5}{\micro s}$ marks the point at which detector is operated.
 (\textbf{d}) Measured (blue dots) and fitted (orange solid line) detector efficiency $\eta_d$ as a function of the duration of the detection step $t_D$. Fit model takes into account bandwidth-limited detection efficiency for short detection windows and $T_1$-decay effect for increasing $t_d$. Black dashed line at $t_D=\SI{5}{\micro s}$ marks the point at which detector is operated, to optimise the photo-detected echo signal.
 }
\label{SupRO}
\end{figure}

\subsubsection{Detector tuning}
In order to perform spin detection the SMPD must be tuned in resonance with the spin-emitted photons.
This is achieved by changing the magnetic flux threading a superconducting quantum interference device (SQUID) embedded in the input resonator of the detector, which allows a tunability range of about \SI{200}{MHz} (see fig.~\ref{SupSMPD1}a and Fig1 of the main text).
The photon detector will be now characterised in the vicinity of this working point.

\subsubsection{Qubit readout}
The qubit readout is based on standard dispersive readout through the buffer resonator. A tone is sent at the buffer resonance frequency pulled by the qubit dispersive shift $\omega_b+\chi_{qb}$ ({thus avoiding to drive the spin resonator at $\omega_0 = \omega_b$}). The qubit state is encoded in the phase of the reflected tone which is subsequently amplified, demodulated and numerically integrated. The readout performances are evaluated by histograming this reflected signal conditioned on the application of a $\pi$ pulse in Fig.~\ref{SupRO}a and b. We observe two Gaussian distributions separated by $10\sigma$ corresponding to the qubit being in the ground and excited state, the spurious tail between the Gaussian corresponds to relaxation events of the qubit during the readout time. A threshold enables the discrimination of the qubit state. It is chosen to minimise the ratio $p(e|0)/p(e|\pi)$ between the false positive and true positive, therefore optimising the dark count rate with respect to the efficiency. We measure a ground state fidelity of $1-p(e|0)=99.2 \% $ and an excited state fidelity of $p(e|\pi)=71 \%$

\subsubsection{Efficiency calibration}

The first characterisation of the device as a photon detector consists in tuning the 4-wave mixing process by sending a weak coherent tone at the detector frequency while scanning the pump frequency and amplitude.
When the 4-wave mixing matching condition is satisfied the qubit is left in its excited state, as shown in Fig.~\ref{SupRO}d.
The chosen working point is the one maximising the probability of the qubit excitation while minimising the residual qubit excitation due to pump heating or spurious process. Note that the frequency drift for the matching condition corresponds to the qubit AC-stark shift induced by the pump tone increasing amplitude.

The detector efficiency is measured by sending a weak coherent tone onto the buffer resonator with $\bar{n}$ photons on average, and measuring the probability $\mathrm{p_e}$ of finding the qubit in its excited state. This measurement is compared to the excited state probability $\mathrm{p_e^{dark}}$ in the absence of incoming pulse for the same detection window. The detector efficiency for a given detection window $t_\mathrm{d}$ is then defined as $\eta_\mathrm{d}=(\mathrm{p_e}-\mathrm{p_e^{dark}})/\bar{n}$.
The photon number $\bar{n}$ is independently calibrated through measurements of qubit dephasing and AC-Stark shift~\cite{gambetta_qubit-photon_2006} taking into account the flux-noise causing extra broadening of the buffer resonance. Fig.~\ref{SupSMPD1}c shows the measured efficiency as a function of the photon frequency in the vicinity of $\omega_b / 2 \pi = \SI{6.946}{GHz}$. The efficiency for the detection window $t_d=5\ \mathrm{\mu s}$ used in the main text is $\eta_\mathrm{d} = 0.53 \pm 0.1$.
We understand quantitatively the infidelity budget of the detector. One source of inefficiency is caused by the readout excited state fidelity which limits the detection efficiency to $0.71$, the other source of infidelity is due to the $T_1$ decay of the qubit during the detection window which limits the efficiency to $0.75$. {The product of these two figures is close to the measured value of $\eta_d=0.53$, showing that qubit relaxation is the dominant limiting factor.}

\subsubsection{Detector bandwidth}
As long as the qubit lies in its ground state, the detector response (Fig.~\ref{SupSMPD1}c) can be modelled by considering that the buffer and waste resonator are coupled with a constant $\mathcal{G}=-\xi_p\sqrt{\chi_{qb}\chi_{qw}}$ due to the 4-wave parametric process involving the qubit, where $\xi_p$ is the pump amplitude in units of square root of photons and $\chi_{qb} (\chi_{qw})$ the dispersive coupling of the buffer (waste) resonator to the qubit.
One can write down the system of coupled equations for the buffer and waste intra-resonator fields $\alpha$ and $\beta$:
\begin{equation}
    \dot{\alpha}= -i\delta_b \alpha -i \mathcal{G}\beta-\frac{\kappa_b}{2}+\sqrt{\kappa_b} \alpha_{in}
\end{equation}

\begin{equation}
    \dot{\beta}= -i\delta_w \beta -i \mathcal{G}^*\alpha-\frac{\kappa_w}{2}+\sqrt{\kappa_w} \beta_{in}
\end{equation}
where $\delta_b$ and $\delta_w$ are the buffer and waste frequencies in the frame rotating  at the probing frequency and $\alpha_{in}$, $\beta_{in}$ are the respective input field amplitudes. Now using the relation between the intra-resonator fields and the input and output flux $\sqrt{\kappa_b}\alpha=\alpha_{in}+\alpha_{out}$, from the equilibrium solution of the coupled system we can extract the transmission coefficient $\left|S_{21}\right|^2=\left|\beta_{out}/\alpha_{in}\right|^2$. Assuming zero input flux on the waste this leads to:
\begin{equation*}
    \left|\mathrm{S_{21}}\right|^2=\left|\frac{2\mathrm{\xi_p}\sqrt{\mathrm{\kappa_b\kappa_w\chi_b\chi_w}}}{-4\delta_\mathrm{b}\delta_\mathrm{w} 
    + 2i\delta_\mathrm{b}\mathrm{\kappa_w} 
    +2i\delta_\mathrm{w}\mathrm{\kappa_b} 
    +\mathrm{\kappa_b\kappa_w} 
    +\mathrm{\chi_b\chi_w \xi_p^2}}\right|^2
\end{equation*}
This expression can be directly related to the detector efficiency when varying the input photon frequency, Figure ~\ref{SupSMPD1}c show a fit of this expression to experimental data with only $\xi_p$ and a scale factor as free parameters. 
From the curve we extract a bandwidth of $\Delta_\mathrm{det}/2\pi \approx \SI{2.1}{MHz}$.

\subsubsection{Dark Counts}

A key figure of merit of the detector is its dark count rate. We characterise this quantity by applying a reset pulse to the qubit through the waste resonator and by keeping the pump tone turned on while no photon pulse is sent to the buffer resonator.  By varying the duration of the pump tone, we observe an increasing excited state population as shown in Fig.\ref{SupRO}c. The residual qubit population rises with a slope of $1.7\ \mathrm{ms}^{-1}$ from an inital value of $0.35\times10^{-2}$. The qubit reaches a finite population of $1.77\times10^{-2}$ after a few characteristic time $T_1=8.1\ \mathrm{\mu s}$. Note that the qubit is initialised well below its thermal population by the reset process. This finite population can be divided in distinct contributions. In the absence of the pump, we measure a residual excited state population of the qubit of $0.81\times 10^{-2}$  tone which is attributed to out-of-equilibrium quasi-particles in the superconducting film \cite{serniak_hot_2018}. By detuning the pump from the matching condition, the heating effect of the pump alone can be evaluated. We measure a negligible rise of the excited population, smaller than $10^{-3}$, compared to the population in absence of pumping. Therefore, most of the excess qubit population $\delta p_e\sim 0.8 \times 10^{-2}$ can be attributed to the finite temperature of the buffer line. Such a finite thermal occupancy $n_\mathrm{th,buffer}$ triggers the detector over its full bandwidth $\Delta_\mathrm{det}$ and leads to dark counts that are integrated over the qubit lifetime $T_1$. The expected rise of qubit population is thus given by $\delta p_e = \eta_d\Delta_\mathrm{det}T_1 n_\mathrm{th,buffer}$ which gives a thermal occupancy of the buffer line of $n_\mathrm{th,buffer} \sim 1.5 \times 10^{-4}$ that corresponds to a residual temperature for the microwave line of $27\ \mathrm{mK}$.

\section{4. Estimation of the number of spins}

\begin{figure}
\includegraphics[width=\columnwidth]{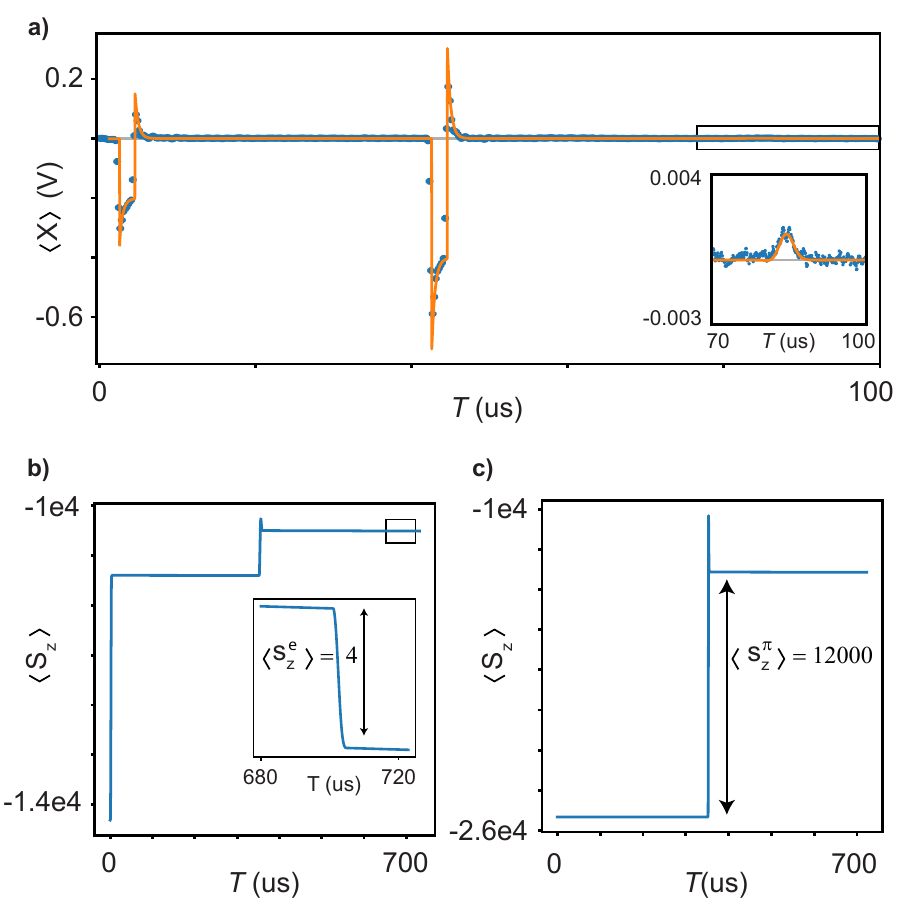}
 \caption{  \textbf{Spin ensemble simulations.}  (\textbf{a}) Measured (blue dots) and simulated (orange solid line) electromagnetic field amplitude at the output of the spin cavity as a function of the time $T$ from the $\pi/2$ pulse of an echo sequence. The echo appears as a slight increase of the field amplitude at twice the separation between the $\pi/2$ and $\pi$ pulse (inset). The spin spectral density $\rho_\mathrm{spin}$ is the only free parameter of the simulation, accordance with the experimental data is achieved for $\rho_\mathrm{spin}=14.6 \ \mathrm{spins \ kHz^{-1}}$.
 (\textbf{b}) Simulated time evolution of $\langle S_\mathrm{z} \rangle$ during an echo sequence, using the same pulse parameters as in the experiment.
 (\textbf{c}) Simulation of the time evolution of  $\langle S_\mathrm{z} \rangle$ during a $\pi$-pulse with the same parameters of the experiment of photo-detected incoherent relaxation. A spin density $\rho_\mathrm{spin}=12 \ \mathrm{spins \ kHz^{-1}}$ was adjusted so that the ratio between the variation of $\langle S_z \rangle$ at echo time (in panel b) and upon the $\pi$ pulse excitation (panel c) reproduces the experimental ratio $C_\mathrm{spin}/(\eta_{duty} c_e)$.
  }
\label{Sup1}
\end{figure}

The goal of this section is to explain how we determine the number of spins $N$ excited in the spontaneous emission detection experiment, enabling us to quantify the overall photon detection efficiency $\eta$.
We use two independent methods, using two different datasets. 

The first method was already used in previous work~\cite{probst_inductive-detection_2017,ranjan_electron_2020}. It relies on measurements performed with homodyne detection, without any use of the SMPD. We measure a complete spin-echo sequence (including the control pulses), and we use simulations to fit the data. The ratio of spin-echo to control pulse amplitude allows fitting the spin density $\rho_\mathrm{spin}$ needed to account for the data as the only adjustable parameter. 

Results are shown in Fig.~\ref{Sup1}a. Quantitative agreement is obtained for $\rho_\mathrm{spin}=14.6 \ \mathrm{spins \ kHz^{-1}}$. We can then determine $N$, the total number of spins excited by a $\pi$ pulse in the spontaneous emission detection experiment of Fig.2 in the main text, by running a second dedicated simulation with spin density $\rho_\mathrm{spin}$. We obtain $N= 14.6 \times 10^3$.

The second method uses a comparison between the two different types of measurements performed with the SMPD. On the one hand, the number of photons $C_\mathrm{spin}$ detected in the spontaneous emission experiment scales linearly with $N$. On the other hand, the echo signal $c_e$ scales like $N^2$, as explained in the main text, indicative of the phase-coherent character of echo emission. Therefore, the ratio $C_\mathrm{spin}/c_e$ can give access to $N$. Note that because both datasets are obtained with the same SMPD and setup, $C_\mathrm{spin}/c_e$ is independent of $\eta$, as required for a proper evaluation of $N$.

To make this reasoning quantitative, we again resort to simulations of both the echo sequence and the $\pi$ pulse. From each simulation, we extract the number of spins involved by computing the change in the total magnetization $\langle S_z \rangle$. The ratio of these two numbers should be equal to the experimentally determined $C_\mathrm{spin}/(\eta_{duty} c_e) $ (the $\eta_{duty}$ correction is due to the fact that echo detection is gated and therefore is insensitive to the detector duty cycle); we use the spin density $\rho_\mathrm{spin}$ as the only adjustable parameter to reach the agreement.

For the experimental value of $C_\mathrm{spin}$ we use the data shown in Fig.~\ref{fig2} of the main text. For $c_e$ we use different pulse parameters than shown in Fig.~\ref{fig3} of the main text (same parameters for the $\pi/2$ pulse, but lower amplitude and $5.5 \mu \mathrm{s}$ duration for the $\pi$ pulse), to get a lower value of $c_e$ and minimize the risk of SMPD saturation. The experimental ratio is then $C_\mathrm{spin}/(\eta_{duty} c_e) = 3 \cdot 10^3$. Figures Fig.~\ref{Sup1}b and \ref{Sup1}c show the evolution of $\langle S_\mathrm{z} \rangle$ at early times in the case of a Hahn echo sequence and of a $\pi$-pulse respectively, with the pulse parameters used in the experiment. The correct ratio is reproduced for $\rho_\mathrm{spin}=12 \ \mathrm{spins \ kHz^{-1}}$. This yields $N = 1.2\times 10^4$.

The two methods are in agreement within an estimated uncertainty $\sigma_\mathrm{spin}\approx 10^3$ on the number of spins taking part to the process. We take the average of the two values $N = (13.3  \pm 1.1)  \times 10^3$ as the reference value for the efficiency estimation. From this, an overall collection efficiency $\eta = 1.1 \times 10^{-2}$ is obtained, as explained in the main text. 

\end{document}